\input epsf.tex
\documentstyle[preprint,prl,aps]{revtex}
\input epsf.tex

\def\sigmav{\mbox{\boldmath$\sigma$}}

\def\vec#1{{\bf #1}}
\def\sigmav{\mbox{\boldmath$\sigma$}}
\def\epsilonv{\mbox{\boldmath$\epsilon$}}
\def\eqn{\begin{equation}}
\def\ee{\end{equation}}
\def\be{\begin{equation}}
\def\frac#1#2{{#1 \over #2}}
\def\endeqn{\end{equation}}
\def\ba{\begin{eqnarray}}
\def\ea{\end{eqnarray}}
\begin{document}
\draft
\preprint{McGill/96-33$~~~$  hep-ph/xxxxx}
\title
{Effective field theories for QED bound states: \\
extending Nonrelativistic QED to study retardation effects
}
\author{Patrick Labelle \thanks{e-mail: labelle@hep.physics.mcgill.ca} }

\address{ Physics Department ,
McGill  University, Montreal , Canada  H3A 2T8}
\date{\today}
\maketitle
\begin{abstract}
Nonrelativistic QED bound states are difficult to study because of the 
presence of at least three widely different scales: the masses,
three-momenta ($p_i$) and kinetic energies ($K_i$) of the constituents.
Nonrelativistic QED (NRQED), an effective field theory developed by
Caswell and Lepage, simplifies greatly bound state calculations by
eliminating the masses as dynamical scales. As we demonstrate, NRQED
diagrams involving only photons of energy $E_\gamma \simeq p_i$
contribute, in any calculation,  to a unique order in $\alpha$.
This is not the case, however, for diagrams involving photons with
energies
$E_\gamma \simeq K_i$ (``retardation effects"),
 for which no simple counting  counting rules
can be given. We present a new effective field theory
 in which the contribution of those
ultra-soft photons can be isolated order by order in $\alpha$. This is
effectively accomplished by performing a multipole expansion of the
NRQED vertices.
\end{abstract}

It is remarkable that the spectrum of the hydrogen atom is one the first
application of quantum mechanics being taught and yet  it is almost
never mentioned  in textbooks on quantum field theory and QED.
Even when the problem of 
  bound states  is mentioned,   it is made clear that it is a
difficult subject and that, to quote Ref.\cite{Itzykson}, ``accurate
predictions require some artistic gifts from the practitioner".

The problem in studying  bound states with relativistic quantum field
theory is that the conventional perturbative
expansion in the number of loops breaks down completely.
 The physical reason is the presence of energy scales absent from
scattering theory. Indeed, the  size of an atom  made of two
particles of charge $-e$ and $Ze$ is of order  
 the Bohr radius $\approx 1 / (Z \mu \alpha)$ (where $\mu $ is the
reduced mass)  which,
by the uncertainty principle, provides an additional energy scale 
 $\approx Z \mu \alpha$. Because
of this new energy scale, there is a region of the momentum
integration in which the addition of  loops will
{\it not} result in additional factors of $\alpha$. Moreover, if $Z \ll
137$ (condition to which we restrict ourselves in this paper),
this energy scale is much smaller than the masses of the particles and the
system is predominantly nonrelativistic, a simplification not
taken advantage of in traditional approaches. In addition, a third
energy scale, again vastly different from the previous two,  is set
by the particles kinetic energies ($\simeq (Z \mu \alpha)^2 / m_i$)
 and further complicates bound state
calculations.

 The problem is greatly simplified by using a Schr\"odinger theory 
 corrected by the usual 
relativistic corrections  obtained by performing a
Foldy-Wouthuysen-Tani \cite{Tani} transformation to the Dirac lagrangian
and expanding   in powers of $\vec p/ m$. These corrections include
   the Darwin interaction, the spin-orbit
coupling, the relativistic corrections to the energy $ (-\vec p^4/(8 m^3)
+ \vec p^6/(16 m^5) + \dots$ and so on
  (from now on, it is this corrected theory that we
will refer to as the ``Schr\"odinger theory").
 The effects of these interactions can be computed
by applying  Rayleigh-Schr\"odinger perturbation theory  using the
Schr\"odinger wavefunctions as unperturbed states,
  a process familiar from
elementary quantum mechanics. Calculations are much simpler in this
framework both because it takes  advantage of the nonrelativistic
nature of the problem and because it sums up the effects of the Coulomb
interaction, responsible for the breakdown of covariant perturbation
theory, into the wavefunctions. 

 However, such a theory is useless for high precision   calculations. 
This is    because
it does not contain the physics corresponding to the high energy ($\vec p
\simeq m$) modes of either the fermions {\it or} the photon.
 This  has two consequences. The first one is the 
 appearance of divergent
expressions. These divergences show up in second order of perturbation
theory (PT) as well as  in first order of PT if sufficiently high order
 (in $1/m$) operators are considered; for example, the operator 
$\vec p^6/ (16 m^5)$ mentioned above
is divergent when evaluated in first order of PT. These
divergences are due to the fact that this theory reproduces
faithfully QED only when the momenta probed by the interactions are
much smaller than the electron mass. This condition is not
satisfied in most interactions considered beyond first order PT, 
 or when the operators contain 
sufficiently high powers of derivatives to probe the relativistic ($p \simeq
 m$)
behavior of the wavefunctions.

The second consequence is the  absence in the Schr\"odinger theory of 
operators corresponding to QED diagrams with photons of
energies $\simeq m_e$, such as the  process
$e^- e^+ \rightarrow \gamma \rightarrow
  e^- e^+$  and the decay of of
an electron-positron pair into an odd or even number of
photons. These processes are clearly
important; the first contributes to the lowest order hyperfine splitting
in positronium, and the others cause the decay of the ortho (total spin
$S$ =1) and para ($S=0$) states of  positronium.

Let us emphasize again that these problems are due to the fact that,
in a quantum field theory, the high
energy modes cannot be simply discarded; they play an important
role, even in processes involving only nonrelativistic external states.

Caswell 
and Lepage (\cite{Lepage}, \cite{KL}) have shown how to modify the
Schr\"odinger theory to  incorporate 
relativistic effects in a consistent and
rigorous manner. They constructed an
   effective field theory (eft) that
reproduces QED in the nonrelativistic regime 
( $\vec p \ll m_e$) and  which they 
christened   nonrelativistic QED (NRQED).
  Although NRQED has been around for more
than ten years and have been used in high precision calculations in 
positronium and muonium (\cite{Lepage},
 \cite{positronium},\cite{Kinoshita}) , it is still little known, both by
the atomic physics community and by the eft afficionados.
Indeed, the Euler-Heisenberg lagrangian,
 which describes the scattering of photons
at energies much below the electron mass, is still the  conventional
example cited as an application of efts in the context of QED.
However,  the Euler-Heisenberg lagrangian (which is a subset of the
NRQED lagrangian) has a range of applications quite limited which
does not include, in contradistinction with NRQED,  the
important topic of bound state physics.

In the next section we will review the construction of NRQED. As any
eft, NRQED  contains an infinite number of interactions and is therefore
nonrenormalizable. This is not a problem because an effective field
 theory is to be
used within a restricted range of energy ($\vec p \ll m_e$ in the
case of NRQED) so that only a finite number of interactions will
contribute to any given process. 
  Which interactions are to be kept for a given precision  (in
$\alpha$) is
dictated by counting rules which are an essential ingredient of any
eft. The counting rules of NRQED are one of   the focus of this paper.

Clearly, NRQED can be applied to both low energy scattering and
nonrelativistic bound states. In applications to bound states,
the NRQED 
  counting rules  are more involved than in most
eft's because of the presence, as noted above, of more than one 
 dynamical 
scale in the theory: the fermions  three-momentum $ \simeq
Z \mu \alpha$, and 
their  kinetic energies $ \simeq
 (Z \mu \alpha)^2/ m_i $.  For the sake of conciseness, from now on
we will refer
to these two scales as, respectively, 
 the ``soft" and ``ultra-soft" energy scales, $E_s$ and $E_{us}$. 
 Because of the  the presence of these two scales, there is, in
general,
no simple connection between an NRQED  diagram and the order  (in $\alpha$)
at which it contributes.

In this paper, we show how to disentangle the contributions
from these two scales in such a way that each  diagram
will contribute to a unique order in $\alpha$. The first step
is well known and relies on   time ordered (or
``old-fashioned") perturbation theory together with the Coulomb gauge
 to separate 
 the  ``soft" photons (with  energy $E_\gamma \simeq E_s$)
 from the ``ultra-soft" ones ($E_\gamma \simeq E_{us}$).  
The counting rules for the diagrams containing only soft photons 
 are straightforward
and  a one-to-one correspondence between a diagram and the
order of its contribution can be established. 
The diagrams with ultra-soft photons 
 are  more complicated; not only do they
contribute to an infinite number of contributions of different order
(in $\alpha$) 
but in addition  the  lowest order
 is not given in terms of simple rules.  This
leads us to the second step in our separation of scales, which 
amounts to performing a multipole expansion of the vertices involving
ultra-soft photons, leading to  a new (infinite) set of interactions.
This can be interpreted as defining a new effective field theory which
is superior to NRQED for dealing with ultra-soft photons and which we
will call ``MQED" (for ``multipole QED"). 
We will show that in MQED,  diagrams containing ultra-soft photons 
 contribute to  a unique    order in $\alpha$, as
given by new counting rules. In addition to having simple counting
rules, MQED is better adapted to the study of processes involving
ultra-soft photons such as the Lamb shift or the generation of certain
types of logarithms. These topics will be addressed elsewhere
\cite{lamb}. 


Our paper is divided as follows. In section I we introduce NRQED and
its Feynman rules, in the context of time ordered perturbation theory.  
In section II we show how time ordered PT permits to separate the contributions
from soft and ultra-soft photons, and give the counting rules for
diagrams containing only soft photons. In section III, we first illustrate
the breakdown of the previous counting rules for diagrams containing
ultra-soft photons. We then  show how to incorporate the multipole
expansion in the NRQED vertices, and how this leads to  defining a new
theory which we will call ``MQED"  for ``Multipole QED". In section IV
we give the general MQED counting rules and some examples to illustrate
their use.
\vfill\eject
\section{NRQED}

In principle, there are two ways of deriving an effective field theory
if the underlying theory is know. Firstly, one can integrate out the
modes of energies $\geq \Lambda_{phys}$ where $\Lambda_{phys}$  is
the energy below which the effective theory is to be used (we
will keep the subscript $phys$ to distinguish this $\Lambda$
from the regulator cutoff to be introduced later on; in NRQED, 
$\Lambda_{phys} \simeq m$). In practice
this is technically difficult to do or even impossible, as  in the
case of low energy QCD. The second method consists in writing down
the most general effective field theory composed of the low energy
fields and consistent with the symmetries of the underlying theory.
The eft is not restricted by renormalizability and contains therefore an
infinite number of operators, each accompanied by an independent
coefficient. If the underlying theory is perturbative in the range of
energy $E \leq \Lambda_{phys}$, then these coefficients can be computed,
order by order in the loop expansion, by setting equal, or ``matching",
some scattering process computed in both the underlying and the
effective theories. In the case of low energy QCD, where such a matching
is not possible, the coefficients must be determined phenomenologically
and the  usefulness of the eft is restricted by the wealth of data
available. 

For NRQED, we follow the second method which requires to first identify
the low energy degrees of freedom and the relevant symmetries.   There
will be a field for the photon and one for each of the charged particles
participating to the process under study such as the electron, the
positron, the muon, proton, etc. Notice that the fermion fields
correspond to two-components Pauli spinors. 
A particle and its 
associated antiparticle are independent fields in a nonrelativistic
field theory; they simply correspond to distinct particles of opposite
charge.  NRQED must obey the symmetries of low energy QED such as
invariance under
parity, Galilean and gauge invariance, etc. Lorentz invariance is not
necessary except for the terms containing  photon fields only.

It is convenient to  decompose the NRQED lagrangian in the following
way:
\ba
{\cal L}_{\rm{NRQED}} ~=~ {\cal L}_{2-Fermi}
 +{\cal L}_{4-Fermi} + {\cal L}_{photon} + \ldots
\ea
where ${\cal L}_{2-Fermi }$ and ${\cal
L}_{4-Fermi}$ are the interactions containing two and four fermions,
, respectively, and 
 ${\cal L}_{photon}$ is the pure photon lagrangian which
includes the Euler-Heisenberg lagrangian. We will not display the operators
containing six or more fermions fields which, in  all practical
applications, can be ignored because their contribution is suppressed. 
 The   lagrangian ${\cal L}_{2-Fermi}$
is given by 
\ba
{\cal L}_{2-Fermi}~ =  & &\psi^{\dagger}\{ iD_{t} + \frac{\vec{D}^{2}}{2m}
+  \frac{\vec{D}^4}{8m^{3}}
+ c_{1} \sigmav \cdot\vec{B}
 \nonumber \\
& &
+ c_{2} (\vec{D}\cdot\vec{E} -\vec{E}\cdot\vec{D})
+ c_{3} \sigmav\cdot(\vec{D}\times \vec{E}
                                - \vec{E}\times \vec{D} )
+ \dots \}\psi    \label{2fermi}
\ea
where  $\psi$  represent a two-component fermion field  of charge $q$
and mass $m$. 
Notice  that in NRQED, a particle and its associated
antiparticle differ only by their charge.

  The first few terms of ${\cal L}_{4-Fermi}$  are given by 
\ba
& c_4& ~  \psi^\dagger \sigmav (-i \sigma_2)  
(\chi^\dagger)^T\,  \cdot\,  \chi^T 
(i \sigma_2) \sigmav
 \psi 
+~c_{5} ~\psi^\dagger (-i \sigma_2) (\chi^\dagger)^T
  \chi^T  (i \sigma_2) \psi
 \nonumber \\ &&~+~ c_{6} \biggl( \psi^\dagger (-i \sigma_2) \sigmav
 \vec D^2 \chi \cdot \chi^\dagger (i \sigma_2)
 \sigmav  \psi +{\rm h.c.} \biggr)
  + ~c_{7}~ \psi^\dagger \sigmav \psi \cdot \chi^\dagger \sigmav
\chi +~ c_{8}~ \psi^\dagger \psi \, \chi^\dagger \chi+ \ldots \label{4fermi}
\ea
The first three terms  are only present when $\psi$ and $\chi$ are
associated with a particle and its antiparticle such as the electron and
positron; they come from QED annihilation diagrams (the factors of
$\sigma_2$ and the transverse operator $T$
 are necessary because we are using the same definition for
both the particle and antiparticle spinors, see (\ref{spinor})).

As will become clear in our derivation of the counting rules, the
Coulomb gauge is the most efficient gauge for the study of
nonrelativistic systems.  In this gauge, the first few terms of $ {\cal
L}_{photon} $ are  \cite{Kinoshita}
\ba
-{1 \over 4 } F_{\mu \nu} F^{\mu \nu} + c_{9} A^0 (\vec k) { \vec k^4
\over m^2} A^0 (\vec k)  - c_{9} A^i(k) {\vec k^4 \over m^2} A^i (k)
~(\delta_{ij} - { k^i k^j \over \vec k^2} ) + \dots \label{photon}
\ea

Before discussing the calculation of the coefficients  $c_i$,
 we will switch from the lagrangian to the hamiltonian. We
 do so 
because the counting rules in a nonrelativistic
bound state  are most easily derived in the context of time ordered
(or  ``old-fashioned") perturbation theory (TOPT for short) and in
TOPT one must work with the hamiltonian rather than the lagrangian. 
We remind the
reader that, in contradistinction with covariant PT, in TOPT
the vertices conserve only three-momenta  and the virtual states
are always on-shell. The total energy, however, is not conserved by the
intermediate state so that,  in this formalism, it is the violation of
energy that characterizes the virtual state rather than  the
off-shellness
of the particles, as in covariant PT.

 Using $\vec D =  i (\vec p - q
\vec A)$ and $D_t = \partial_t + i q A_0$,  the
NRQED hamiltonian is given by 
\ba
&& {\cal H}_{2-Fermi}~ =  \psi^{\dagger} \biggl[ {\vec p^2 \over 2 m} +q A_0
-  { \vec p^4 \over 8 m^3} - { q \over 2m} ( \vec p' + \vec p )
\cdot
\vec A + { q ^2 \over 2m } \vec A \cdot \vec A 
 \nonumber \\ &&~~~~~~~~~~~~~~~~~~~~~~~~~
-  i  c_1
\sigmav \cdot ( \vec k \times \vec A) -  c_2
\vec k^2
A^0 \nonumber   \\ &&~~~~~~~~~~~~~~~~~~~~~~~~~
 +   2 c_3  \sigmav \cdot (\vec p' \times \vec p
) A^0
-  2 q  c_3 \sigmav \cdot (
\vec k_1 \times \vec A (k_1)
) A^0(k_2)  \nonumber  \\ &&~~~~~~~~~~~~~~~~~~~~~~~~~
 +  c_3  k^0 \sigmav \cdot ( (\vec
p' + \vec p) \times \vec A ) + \dots \biggr] \psi (\vec p) ~+~
  \chi^\dagger \chi   {\rm 
 ~terms} \label{H2Fermi} \\ 
&& {\cal H}_{4-Fermi}~ = 
- c_4 ~ \psi^\dagger \sigmav (-i \sigma_2)  
(\chi^\dagger)^T \, \cdot\,  \chi^T 
(i \sigma_2) \sigmav
 \psi 
~-~c_{5}~ \psi^\dagger (-i \sigma_2) (\chi^\dagger)^T \,
  \chi^T  (i \sigma_2) \psi + \dots
 \label{H4Fermi}
 \\&& {\cal H}_{photon}~ =  {1 \over 2} \bigl( \vec E^2 + \vec B^2 \bigr) 
-c_{9} A^0 (\vec k) { \vec k^4
\over m^2} A^0 (\vec k)  + c_{9} A^i(k) {\vec k^4 \over m^2} A^i (k)
~(\delta_{ij} - { k^i k^j \over \vec k^2} ) + \dots \label{hamiltonian}
\ea

As explained previously, the coefficients are determined by computing
some low energy scattering process
 in both QED and NRQED and matching the results. The
coefficients of the operators  in ${\cal H}_{2-Fermi}$
 can be computed by considering the scattering of a charged particle off
an external field (see Ref. \cite{Kinoshita} or  \cite{thesis}
 for an explicit matching). 
 The coefficient $c_4$ is obtained
by matching  the tree level QED annihilation diagram $e^+ e^- 
\rightarrow \gamma  \rightarrow e^+ e^-$ to the NRQED interaction.
The tree  level  contribution to $c_5$  comes from the QED diagram
$e^+ e^-
\rightarrow \gamma  \gamma  \rightarrow e^+ e^-$ and is therefore of
order $\alpha^2$. On the other hand, $c_9$ comes from the one-loop
vacuum polarization.  One finds
\ba
c_1 &=& { q \over 2 m} ~~~~~~~~~~
c_2 = { q \over 8 m^2} \nonumber \\
c_3 &=& {i q \over 8 m ^2}  ~~~~~~~~~~
c_4   = - { \alpha \pi \over m^2} \nonumber \\
c_5  &=&{  \alpha^2 \over m ^2}  ( 2 - 2 \ln 2 + i \pi )  ~~~
 c_9 = {\alpha \over 15 \pi}  .
\ea
The imaginary part of $c_5$ corresponds, via the relation ${\rm Im (E)}
=- 
\Gamma /2$,  to the decay rate of positronium
in a singlet ($S=0$) state,  the quantum number carried by the
corresponding operator.

 Notice that the relation between the powers of
$\alpha$ and the number of loops is broken in NRQED, since factors of
the coupling constant arise from coupling to all photons whereas the eft
contain only photons with momenta $\vert \vec  k \vert 
 \ll m$. For the same reason, the
tree level matching, for example,  involves  NRQED  tree diagrams, but
may involve   QED loop diagrams. By ``tree level matching" we will mean
matching involving tree level NRQED diagrams.

  The one-loop matching 
 modifies  the values of the tree level  coefficients so that we will,
from now on, write the coefficients in the form
\ba
c_i \rightarrow c_i \delta_i
\ea
with $\delta_i = 1 + {\cal O } ( \alpha)$. As in conventional
renormalization, 
 tree level as well as one-loop NRQED   diagrams 
 enter in the one-loop matching
 and this defines the ${\cal O} ( \alpha)$ corrections to the 
NRQED parameters; the only difference with conventional renormalization
is that the calculation is matched
to a QED result instead of an experimental input. 
Because the one-loop NRQED integrals are divergent, they must be
regularized. There are many possible regulators; one can use
dimensional regularization or a simple cutoff $\Lambda_R$ on the momentum
integrations (which is permitted  because
NRQED breaks Lorentz invariance to start with). The NRQED coefficients
defined by the matching are then cutoff dependent, {\it i.e.} they
must be viewed as bare parameters. In contradistinction with
QED, the divergent terms are not only logarithmic  but power-law, $(\Lambda_R/
 m)^n$, as well.
 This  cutoff dependence is of course
canceled in any physical calculation, by  invariance under the
renormalization group. Obviously, one can also set $\Lambda_R =
\Lambda_{phys} = m $ directly, but since the bare coefficients are then
finite,  this can be misleading if one is
not careful about renormalizing the effective theory properly (for a 
more thorough discussion of this point, see \cite{Cliff}).

The one-loop matching of some of the coefficients appearing in (\ref{2fermi})
has been performed in Refs. \cite{Kinoshita} and  \cite{MRST}
 and the corresponding $\delta_i$'s appearing in
(\ref{2fermi}) were found to be 
\ba
&& \delta_1 \equiv  \delta_F ~=~ 1 +   a_e + {\cal O} ( \alpha^2) 
\nonumber \\ \nonumber
&& \delta_2 \equiv \delta_D
  ~=~ 1 + {\alpha \over \pi} {8 \over 3} \biggl[ \ln \bigr( { m
\over 2 \Lambda_R} \bigr) + {11 \over 24} \biggr] + 2 a_e   + {\cal O} (
\alpha^2)   \\
\nonumber
&& \delta_{3} \equiv \delta_{S} ~=~ 1 + 2 a_e
 + {\cal O} ( \alpha^2) \nonumber \\ &&
\delta_{4} \equiv \delta_{A} ~=~ 1 - {44 \alpha \over 9 \pi} \label{renor}
\ea
where $a_e$ is the electron anomalous magnetic moment which,
to the order of interest, can be taken to be $ \alpha/ (2 \pi)$.   We have
redefine our coefficients to follow the convention of \cite{Kinoshita}
(but notice that  our $\delta$ correspond to their coefficients $c$);
 the subscripts $F, \, D , \, S$  and $4-F$ stand for Fermi,
 Darwin, spin-orbit and four-Fermi   interaction, respectively.

We now turn to the task of writing a general form for the NRQED
coefficients. Before doing so, we
must address the issue of the photon mass, which provides an additional 
scale and has the potential of complicating our analysis. The photon
mass does not appear in (\ref{renor}) but this might appear fortuitous.
However, since
any photon mass dependence is a sign of sensitivity to very low momenta
and NRQED is designed to be equivalent to QED in this region of phase
space,   any infrared singularity in a  QED  diagram is
also present in the corresponding NRQED diagram, so that it gets
canceled in the matching.
Therefore, in general,
the NRQED bare coefficients do not depend on the photon mass,
to any order in the matching. From this, it follows that the
coefficients have the general
structure
\ba
c_i (\Lambda_R,m_1,m_2)~ =~ c_i^0~ \alpha^{n_i}~\delta_i (\Lambda_R,m_1,m_2)
\equiv c_i^0~ \alpha^{n_i}~ \bigl[
 1 +~~ \sum_{l_i =1}^\infty~ \alpha^{l_i}~\,  \tilde  
c_i^{l_i}(\Lambda_R,m_1,m_2)~
 \bigr]  
   \label{coeff}
\ea
where $c_i$ is now a generic symbol representing any NRQED coefficient
and the $l_i$ on the coefficients $\tilde c_i$ is an index, not an  exponent.
We have  decomposed the lowest order term as a coefficient $c_i^0$ of
order one times a factor $\alpha^{n_i}$ which is different for different
operators.
As an example, 
 the  Darwin interaction, which contains the factor $c_1$, has 
$n_1= 1/2$ whereas the singlet annihilation operator, which contains
$c_5$, has $n_5=2$. 
The index  $l_i$
indicates the  number of  loops  
 used in the matching. 

The coefficients $  \tilde c$
 contain, in general,  finite pieces plus  power-law
terms as well as logarithms divergent terms.
  Notice that for a fixed $l_i$, there are,
in principle,  an
infinite number of terms  to calculate because there are an infinite
number of $l_i-  {\rm   loops} $   NRQED Feynman diagrams but 
 only a finite number of
interactions must be considered in any given calculation, as specified
by the counting rules (which will also dictate the order at which the
matching must be performed).

%

The Feynman rules for the first few interactions of
Eqs.(\ref{hamiltonian}), (\ref{H2Fermi}) and  (\ref{H4Fermi}) 
are given in Fig.[1]. 
  We will draw the diagrams with the time
flowing to the right. In the rules for the vertices we have followed the
example of  \cite{Kinoshita} and used the expression
 ``dipole vertex" to represent
 the $\vec p \cdot \vec A$ interaction even though,  as pointed out in
\cite{Kinoshita}, the NRQED hamiltonian is not an expansion in
multipoles. Also, we have use some Fierz  reshuffling to rewrite the
annihilation vertex in the form given in Fig.[1].
As for the propagators, we have used 
    time ordered perturbation theory where there is 
one propagator for each different  intermediate state,  defined by
cutting the diagram with a vertical line. The  general rule for a
time-ordered 
propagator is
\ba 
{1 \over E_0 - E_i }  \label{prop}
\ea
times a factor
\ba
{1 \over 2 k} ~ \bigl( \delta_{ij} - { k_i k_j \over \vec k^2 +
\lambda^2} \bigr)
\ea
for each transverse photon present in the intermediate state. In 
(\ref{prop}), $E_0$ stands for the energy of the initial state and $E_i$
for the energy of the intermediate state. One uses
nonrelativistic energies, $\vec p^2 / (2 m)$, for the fermions  and 
$\sqrt{ \vec k^2 + \lambda^2}$ for the photons. In Fig.[1] the
propagator is given for an intermediate state containing only one
fermion or one transverse photon. In Fig.[2],  the corresponding
expressions are given for the states containing two fermions or two
fermions plus one transverse photons, which are the situations most
often met in NRQED calculations.

  One must sum over all the possible time ordered
diagrams and integrate over all the internal three-momenta, with a
measure $d^3p /(2 \pi)^3$.  Notice that we prefer to include the factors
of $1/(2 \sqrt{\vec k^2 + \lambda^2})$ 
corresponding to the transverse photons in the propagators
instead than in the measure for reasons that will become clearer below.

In this work we  will be mainly interested in applications of NRQED to
bound state calculations  in which case 
 the external lines are not
associated with free spinors but with wavefunctions. In general, the
wavefunctions are obtained by solving a Bethe-Salpeter type
equation, with some approximated kernel.  This is equivalent to summing
up an infinite number of this kernel into the wavefunctions. We will
show below that the NRQED counting rules single out (in the Coulomb
gauge) the Coulomb interaction as being the only nonperturbative
interaction in a nonrelativistic bound state so that this 
part of the analysis reduces to solving the  usual Schr\"odinger equation.
In our explicit examples, we will use the ground state wavefunction, given
by
\ba
 \Psi(\vec p)_{n, l=0,s1,s2} ~=~
 { 8 \sqrt{ \pi \gamma^5} \over ( \vec p^2 + \gamma^2)^2}
\otimes \xi_1 \otimes \xi_2 \label{wave}
\ea
where $ \gamma \equiv  Z \mu \alpha  $ (the energy of the state is given
by $- \gamma^2 / (2 \mu)$ )  and $\xi_1, \xi_2$ are 
 the spinors of the two particles making up the bound state, 
with 
\ba
\xi_{ up} = \pmatrix{1 \cr 0},~~\xi_{ down} = \pmatrix{0
\cr 1} . \label{spinor}
\ea

We will not write down the states of higher angular momentum since they
are, for the purposes of establishing the counting rules, equivalent to
the above states (for the momentum Schr\"odinger wavefunctions for
arbitrary quantum numbers, see \cite{moment}).  As just mentioned,
 using Schr\"odinger wavefunctions for the external
states means that we are summing the Coulomb interaction between the
external legs. All other interactions can be treated perturbatively,
which will  be shown to be self consistent 
 with the counting rules.

\section{Counting rules: soft photons}

We now  consider a nonrelativistic 
  bound state made, to simplify the discussion,  of two particles of equal
masses and of charges $\pm e$. We will also assume that it is 
in its ground state ($n=1$).
  We will generalize our results at
 the end of this section.

There are two important energy scales in such a bound state,
the typical bound state momentum $ \gamma$ and the binding energy $-
\gamma^2/m$.  For a nonrelativistic fermion, for which the dispersion
relation is given by the usual $E= \vec p^2/(2m)$, using either scale
leads to $p_{fermion} \approx \gamma$ (from now on, by $p$ and $k$ we
will always mean the magnitude of three-momenta). In the case of a photon, for
which $E= k$, using the bound state momentum or binding energy yields
two very different scales for $k$, namely $k\simeq \gamma= m \alpha$
and $k \simeq \gamma^2/m = m \alpha^2$. We will refer to this first
type
of photons as ``soft" photons, and to the second type as ``ultra-soft"
photons. For the sake of completeness, we define ``hard photons" as
the photons with $k \simeq m$ or greater. These photons play no
dynamical role in NRQED since they have been integrated out of the
theory and their only effect is buried in the theory's coefficients.

The first step in deriving counting rules is to separate diagrams
involving soft photons from diagrams with ultra-soft photons, since
they bring in very different scales, which will necessarily complicate
the rules. This is where the use of the Coulomb gauge in conjunction
with time ordered PT will be crucial in simplifying the analysis.

Consider  a transverse 
photon exchange   between two  fermions in a nonrelativistic  bound state.
This is represented by the two time ordered diagrams of Fig.[3], where we
put the time axis toward the right and  the $\Psi$ attached to the
external lines represent the wavefunctions.
 The photon will contain  both  soft and  ultra-soft components. Now, if the
photon is soft, its momentum as well as its energy  are 
 of order the fermion momentum $ m  \alpha$  so that its
energy is much greater than the fermion energies. 
This means that from the point of view of the fermions, the propagation
of the soft photons is instantaneous and is therefore represented by
vertical lines in time-ordered diagrams.

This can be seen more qualitatively by looking at the explicit
expression for the intermediate state propagator, which is given by
(recall that $k \equiv \sqrt{ \vert \vec k^2 \vert} $)
\ba
  \, { 1\over 2 k} \, \bigl( 
\delta_{ij}   - { k_i k_j \over \vec k^2 } \bigr) &&
    \biggl( {1 \over 
-\gamma^2/(m)  -
\vec p^2/(2 m) - (\vec p - \vec k)^2/(2 m) - k
} \nonumber \\ ~&+&~ {1 \over  -
\gamma^2/( m)  -
\vec p^2/(2 m) - (\vec p - \vec k)^2/(2 m) -  k
 } \biggr)
. \label{tere}
\ea
Notice that the photon mass can
 be set to zero in bound states calculations, the
size of the atom preventing the appearance of any infrared singularity;
the scale of the fermion three-momentum $\vec p$  is of order $\gamma$.
 For soft photons, $  k \simeq Z \mu \alpha$, 
and we
clearly see that $k$ dominates in the propagators for the intermediate state
so that we can approximate (\ref{tere})  by
\ba
 \, { 1\over 2 k} \, \bigl(
\delta_{ij} - { k_i k_j \over \vec k^2 } \bigr)
\,\biggl( { - 2 \over k}
\biggr)
~=~ 
 -{   1 \over \vec k^2}  \bigl(
\delta_{ij} - { k_i k_j \over \vec k^2 + \lambda^2} \bigr) \label{softphoton}
\ea
which  corresponds to a single diagram, with an energy
independent  photon propagator. This corresponds to the transverse
photon propagator of \cite{Kinoshita} if one approximates $k_0^2 - \vec
k^2 \approx
- \vec k^2$ and set $\lambda=0$ (this is why we kept the $1/(2 k)$
factor in the definition of the propagator instead of the measure).
 This shows again that
in a time ordered diagram the
propagation of such a photon is represented by a vertical line, {\it
i.e.} an instantaneous interaction,  since it is independent of $k_0$
so that its Fourier transform contains a delta function in time.

We can now isolate the soft from the ultra-soft components in any photon
exchange by rewriting the time ordered diagram as a sum over an
instantaneous interaction and a ``retarded" one, as in Fig.[4] (this is
why we refer to the effects of ultra-soft photons as ``retardation
effects"). If we
restrict ourselves to NRQED diagrams containing soft photons only, then
all photon exchanges can be  represented by
vertical lines.
 In real space, such interactions are
represented by potentials local in time, {\it i.e.} functions of $\vert
\vec r_1 - \vec r_2 \vert $ only.

Besides photon exchanges, the only other possible interactions are the
self energy interactions such as $-p^4 / 8 m ^3$ and the contact
interactions contained in ${\cal L}_{4-fermi}$, ${\cal L}_{6-Fermi}$,
etc. These can clearly be represented by potentials, so that an
arbitrary diagram containing soft photons only can be written as a
string of potentials connected by fermion lines only. 
In this case, the  intermediate states contain fermion
lines  only and  the time ordered propagators take on a particularly
simple form.  If there are no interaction from ${\cal L}_{n>4}$, for
example, the propagators are all of the form
\ba
{1 \over E_0  - E (intermediate ~state) } ~=~
{1 \over -\gamma^2/m - \vec p^2/(2 m) - \vec p^2/ (2 m)}
~=~ - {  m \over \gamma^2 + \vec p^2}
\ea
where we have used the fact that the bound state energy is $-
\gamma^2/ m$. The crucial observation is that the mass dependence
factors out, in the form of the overall factor of $m$, leaving
$\gamma$ as the only dynamical scale in the integrals.
If interactions contained in ${\cal L}_{n>4}$ are included, then some
intermediate states will contain more than two fermion lines, but it
will always be of the form 
\ba
{1 \over -\gamma^2/m - \sum_i \vec p_i^2/(2 m) }~=~ - {  m \over
\gamma^2 + \sum_i \vec p_i^2}
\ea
and the mass still factors out.
  Therefore any
NRQED diagram containing only soft photons leads to an integral of the
form
\ba
m^b ~\biggl( \Pi_j  c_j(\Lambda_R)  \biggr)  \int^{\Lambda_R}
\bigl( \Pi_i  d^3 p_i \bigr)
~F(\vec p_i, \gamma) \label{softintegral} 
\ea
where $b$ is some integer that depends on the types and number of
potentials.
 The product is over all the vertices of type
$j$, with
coefficients $c_j$, as given in (\ref{coeff}). Again, the crucial point for
the following discussion is that the mass $m$ does not appear in the
integrand, {\it i.e.} does not play any dynamical role.
 There are two scales in
the integral, $\gamma$ and $\Lambda_R$,  but
the invariance under the renormalization group  implies that the 
divergent
$\Lambda_R $ dependent terms  arising from the integrations will be
canceled by  corresponding  terms in the bare coefficients
$c_i(\Lambda_R)$. As noted before, these
 divergent terms are either power-law, {\it i.e.}
of the form $(\Lambda_R)^n$ with n being a positive integer, or
logarithmic.
The power law terms are canceled exactly whereas
the $\Lambda_R$ in the logarithms get canceled after combining
logarithms containing different scales which leaves,  in  the end, 
 logarithms of $\alpha$.

 How the logarithms become finite  is instructive in
that it clearly illustrates the separation of scales accomplished by the
effective theory. As mentioned in section 2,  some NRQED bare
coefficients contain divergent logarithms of the form
 $\ln (\Lambda_R/m)$ (as is explicit in
(\ref{coeff})). To be precise,
the NRQED scattering diagrams appearing in the matching process contain
logarithms of $\Lambda_R$ over $m$ 
 since these are the only two dynamical scales of the
eft, whereas the QED scattering diagrams  contain logs of the form
$\ln (m/ \lambda)$; upon solving for the bare coefficients, the
logarithmic dependence is then of the form $\ln (\Lambda_R / m)$ (again,
the photon mass dependence drops out entirely for the reasons explained
above).  On the other hand, the NRQED bound state integrals can only
depend on the scales $\Lambda_R $ and $\gamma$, yielding $\ln (
\Lambda_R / \gamma)$. In the end, the  logs of the bare coefficients
combine with the logs generated by the bound state integrals to give
corrections of the form
 $\ln ( \gamma/ m) = \ln \alpha$. 
We see how the use of an effective theory has separated the
contributions from all the scales present in the problem ($\lambda,
\gamma, m $~and~$ \Lambda_R$) in such a way that only two of them played a
dynamical role in any given stage of the calculation (QED and NRQED
scattering diagrams in the matching and NRQED bound state diagrams).

The only $\Lambda_R$  dependence remaining is therefore of the form 
$(\gamma / \Lambda_R)^n$ which, upon setting $\Lambda_R = m$, leads to
corrections beyond the order of interest; in analytical calculations one
can get rid of these terms by simply   letting
$\Lambda_R \rightarrow \infty$ at the end of the calculation, as in
conventional renormalization. 
In numerical calculations, one needs to keep $\Lambda_R$ finite
because intermediate steps of the calculations are divergent (unless
bound state diagrams and counterterms are combined before carrying the
integals);
one must then restrict $\Lambda_R$ to the range $\gamma \ll \Lambda_R
\ll m $ in order to assure convergence of the expansion in $1/m$ used in
the effective theory. In this case
  one must keep track of the finite corrections due to the
finite value of the cutoff\footnote{Obviously,
 dimensional regularization can be used instead of a momentum
cutoff. The power law divergences are then either entirely absent or
replaced by $1/ \epsilon$ divergences. Again, these divergences cancel,
by invariance under the renormalization group. This leaves logarithms
depending on the scale $\mu$, which gets canceled in the way described
above for the $\log \Lambda_R$ terms. In actual explicit analytical 
calculations, using dimensional regularization or a momentum cutoff is
simply a matter of taste. However, for high precision calculations,
where numerical calculations are required, an explicit cutoff is necessary.}.

It is now a trivial matter to write down the counting rules for an
arbitrary bound state diagram containing only soft photons,
 {\it i.e.} the order in $\alpha$ at which
it will contribute. There are
two sources of factors of $\alpha$.
First, there are the explicit factors contained in the NRQED vertices.
Secondly, there is a factor of $\alpha$ for each factor of $\gamma$
generated by the diagram. To be more rigorous, the factors associated to
the vertices are genuine factors of the coupling constant whereas the
factors of $\gamma$ are associated with factors of $v$ which scale is
set by the bound state to be of order $\alpha$; here it is not important
to distinguish between the two types of contributions but this is
necessary in QCD bound states because of the noticeable running of the
strong coupling constant \cite{NRQCD}, \cite{Pat}.

 By
simple dimensional analysis,  there will be a factor of $\gamma$ to
compensate each explicit factor of mass appearing in the vertices and
each factor of mass due to the fermion pair time ordered propagators.
An arbitrary bound state diagram  is built out of a given number of
potentials, $N_{{\cal
P}}$, 
connected  by $N_{TOP}$ time ordered propagators. For example, consider
Fig.[5], where $3$ potentials are connected by $2$ time-ordered
propagators so that $N_{{\cal
P}} =3 $ and $ N_{TOP} =2 $ for  that diagram. For later use, we also
define ${\cal V}_i$ as the number of vertices contained in the $i^{th}$
potential and ${\cal V}$ as the total number of vertices in the diagram
\ba
{\cal V} = \sum_{i=1}^{N_{{\cal P}}} {\cal V}_i
.
\ea
We now define the ``vertex mass degree" $d_j$ as the number of {\it
inverse} masses contained in the $j^{th}$ vertex and the ``potential
mass degree" 
  ${\cal D}_i$ as the number of inverse masses contained
in the $i^{th}$ potential,
\ba
{\cal D}_i = \sum_{j=1}^{N_{{\cal V}_i}}  d_j .
\ea
  For example, the first potential
of Fig.[5] (the self energy potential) has ${\cal D}_1 = 3$,
whereas ${\cal D}_2 =0$  (for the Coulomb interaction) and ${\cal D}_3
= 4$. Since each potential generates ${\cal D}_i$ factors  of inverses
masses and each fermion-fermion time ordered propagator generates one
factor of $m$, 
an arbitrary diagram having the dimensions of energy 
will then generate a factor $\gamma^\lambda$, with 
\ba
\lambda = 1   - N_{TOP} + \sum_{i=1}^{{\cal V}}  d_i \label{peace}
\ea
where the sum is over all the vertices in the diagram. For the present
purposes, it is more convenient to write $\lambda$ as
\ba 
\lambda = 1   - N_{TOP} + \sum_{i=1}^{N_{{\cal P}}}  {\cal D}_i 
\label{lambda}
\ea
where now the sum is over all the potentials in the diagram.

We know define the ``coupling constant degree" ${\cal C}_i$ as the
total number of explicit factors of $\alpha$ contained in each
potential, namely
\ba
{\cal C}_i \equiv \sum_j^{{\cal V}_i}  (n_j + l_j) \label{Ci}
\ea
where 
 $n_j$ and $l_j$ are the
powers of $\alpha$ associated with the coefficient of each vertex
 as defined in (\ref{coeff}).

Finally, a  diagram made
 of $N_{{\cal P}}$ potentials  will contribute to order $m
\alpha^\zeta$ with $\zeta$ being the sum of  Eq.(\ref{lambda})
and  the coupling constant degrees (\ref{Ci}) of all the potentials:
\ba
\zeta = \sum_{i=1}^{N_{{\cal P}}} \bigl(
 {\cal D}_i + {\cal C}_i \bigr) +1- N_{TOP} . \label{deltaone}
\ea
It is easy to see that $N_{TOP}$ and $N_{{\cal P}}$ are related by
$N_{TOP} = N_{{\cal P}} -1 $ so that we can write 
\ba
\zeta = \sum_{i=1}^{N_{{\cal P}}}  \bigl({\cal D}_i + {\cal C}_i
  \bigr) +2-
 N_{{\cal P}} ~=~ 
 \sum_{i=1}^{N_{{\cal P}}} \bigl(
{\cal D}_i + {\cal C}_i
 - 1 \bigr) +2 \label{delta}.
\ea
Ths expression  gives the order in $\alpha$ of any NRQED diagram
containing only soft photons, 
keeping in mind that this result can be enhanced by factors of
$\ln(\alpha)$. For  the example of Fig.[5], one finds $\zeta =8$. 

Eq.(\ref{delta}) shows clearly that if there is a potential for which $
{\cal D}_i + {\cal C}_i
 =1$,
perturbation theory will break down and it will have to be summed up to
infinity. It is an easy matter to find such a potential.
 We can choose $l_j=0$ ({\it i.e.} the coefficients of the
vertices have their tree level values). Now,  $n_j$ is zero for the
self-energy interactions, but the lowest value that  the mass degree
  can take is
$3$, corresponding to the interaction   $-p^4 / (8 m^3)$, so that
the condition $ {\cal D}_i + {\cal C}_i
 =1$ cannot be fulfilled.
Many potentials have $n_j=1$ ({\it i.e.} one factor of $\alpha$) but the
only one with, in addition,   ${\cal D}_i =0 $ (no inverse masses)
 is the Coulomb potential $-e^2 /\vec
k^2$. Therefore,
 as expected, only the Coulomb interaction must be summed up to
infinity and the resulting contribution is, from (\ref{delta}),
of order 
 $m \alpha^2$; all other potentials can be treated in perturbation
theory.

In an actual calculation, the counting rules are used in the following
way. For a given process (hyperfine splitting, decay rate, etc.),
 one  selects all the diagrams with the appropriate quantum numbers that
will, using the counting rules, contribute to the order 
of interest.  The counting rules determine not only the diagrams that
must be retained, but also, via the $l_j$ dependence in (\ref{delta}),
the number of loops that must be used in the matching of each vertex.
The matching is then carried for each vertex using scattering diagrams
in
both QED and NRQED. For a given number of loops, there are an infinite
number of NRQED scattering diagrams, but here the  counting rules are used 
a second time to pick the NRQED diagrams that need to be taken
into account. Notice that 
 in the matching process, which involves scattering
diagrams, one uses (\ref{delta}) even though this relation was derived
for bound state diagrams. Once all the relevant diagrams have been taken
into account and the NRQED coefficients have been renormalized to the
appropriate order, the final calculation will be finite and will
reproduce the QED result, to the order of interest.

\subsection{Extension to arbitrary masses and charges}

We now extend our counting rules for two constituents having
arbitrary masses $m_1$ and $m_2$. The above derivation must then be
modified at two points. First, the NRQED coefficients given by
(\ref{coeff}) will now contain a dependence on $m_1$ and $m_2$:
\ba
c_i (\Lambda_R, m_1,m_2)~ =~ c_i^0(m_1, m_2)~ \alpha^{n_i}~\Biggl[
 1 +~~ \sum_{l_i=1}^\infty~ \alpha^{l_i}~\,
  \tilde  c_i^{l_i}(\Lambda_R,m_1,m_2)~
 \Biggr]  
  . \label{coeff2}
\ea
No simple general expression can be given for the mass 
  dependence  of the coefficients $\tilde c$; it arises from 
QED loop diagrams entering the matching and may involve logarithms of
$m_1 / m_2$, etc. The mass dependence of the zeroth order coefficients
$c_i^0$ can be taken into account in the following way:
 first, define the vertex mass degrees with respect to each mass,
  $d_j(m_1)$ and $d_j(m_2)$  as the number of inverse masses $m_1$ and
$m_2$ contained in the vertex.
For a given NRQED bound state
 diagram, one can then define the following two indices
\ba
\kappa \equiv \sum_{i=1}^{\cal V} d_i(m_1), \\
\rho \equiv \sum_{i=1}^{\cal V} d_i(m_2) .  \label{kappa}
\ea

Obviously,  such a diagram will contribute an overall factor
$1/(m_1^\kappa m_2^\rho)$. Since the overall result must have the
dimensions of energy and since the only energy scale provided by the
bound state NRQED diagrams is $ \gamma$, which contains the reduced
mass $\mu$, the overall mass factor will be
\ba
{\mu^{  \kappa+ \rho +1} \over m_1^\kappa m_2^\rho} 
.
\ea

The more general counting rule is therefore
\ba
{\cal O} =   {\mu^{\kappa+ \rho +1} \over m_1^\kappa m_2^\rho}
 \alpha^\zeta \label{order1}
\ea
times possible factors of $\ln  \mu \alpha$ and functions of the masses
 $m_1, \, m_2$  and $\mu$ (which, however, arise only if some of the NRQED
coefficients have been matched beyond tree level).

Finally, we consider  a bound state with constituents of charges
$ - e$ and  $ Z e$.
 We first include a $Z$ dependence in the NRQED
coefficients:
\ba
c_i (\Lambda_R, m_1, m_2, )~ =~ c_i^0(m_1, m_2)~Z^{a_i}~ \alpha^{n_i}~\Biggl[
 1 +~~ \sum_{l_i
 = 1}^\infty~ \alpha^{l_i}~\,  \tilde  c_i^{l_i}(\Lambda_R,m_1,m_2,Z)~
 \Biggr]  
   \label{coeff3}
\ea
where $a_i$ will denote the explicit power of $Z$ contained in the
zeroth order coefficient of the 
$i^{th}$ vertex.
Again, the  $Z$ dependence of the $ \tilde c_i^{l_i}$
 arises from the computation of
QED loop diagrams and we will not write a general expression
 for this dependence but notice that it will necessarily be some power of
$Z$. There is an additional $Z$ dependence which, this time, we can
take into account:  an additional  $Z$ dependence
comes from each factor of $\gamma = Z  \mu \alpha $
 generated by the NRQED bound states. This number is given by
 (\ref{peace}):
\ba
\lambda = 1   - N_{TOP} + \sum_{i=1}^{{\cal V}}  d_i. 
\ea
 A bound state diagram (with all the NRQED coefficients  taking their
tree level value)  will therefore generate a factor $Z^\eta$ with
$\eta$ given by this last
expression plus the $Z$ dependence of the tree level NRQED coefficients,
as given  in (\ref{coeff3}):
\ba
\eta = 1   - N_{TOP} + \sum_{i=1}^{{\cal V}} \bigl(  d_i + a_i \bigr). 
\label{eta}
\ea
Again, the power of $Z$ is independent of the order in perturbation
theory for the Coulomb interaction since each Coulomb potential
increases the sum over $a_i$ by one but increases $N_{TOP}$ by one as
well.
Our  most general counting rule for diagram containing soft photons
 is  therefore given by
\ba
{\cal O} =   {\mu^{\kappa+ \rho +1 } \over m_1^\kappa m_2^\rho} Z^\eta
 \alpha^\zeta \label{order2}
\ea
times possible factors of $\ln(Z  \mu \alpha)$, and dependence on $m_1, \,
m_2, \, \mu $
and $Z$ arising from  the loop corrections to the NRQED coefficients.

\section{Counting rules: ultra-soft photons}

The above derivation relied heavily on the fact that the only scale
present in the bound state diagram was $\gamma$. However, if we start
considering ultra-soft transverse
photons, then we have to go back to the general 
time ordered propagator (see Fig.[3])
\ba
 -  {{\cal P}_{ij} \over 2 k}&&  \biggl( {1 \over 
\gamma^2/(2 \mu)  +
 (\vec p - \vec k)^2/(2 m_1) 
+
\vec p^2/(2 m_2) +k } \nonumber \\ &&~~~~~~~~~~
~+~ {1 \over  
\gamma^2/( 2 \mu)  +
\vec p^2/(2 m_1) + (\vec p - \vec k)^2/(2 m_2) +k } \biggr)  
\label{generalpropagator}
\ea
where we have defined the transverse projection operator
\ba
{\cal P}_{ij} \equiv \delta_{ij} - { k_i k_j \over \vec k^2} .
\ea
In general, such a propagator would 
 contain both  the  soft and ultra-soft scales so that counting rules would
be impossible to establish. However, we have already isolated the soft
contribution in an instantaneous interaction with the photon propagator
given by (\ref{softphoton}). Therefore, if the contribution from the
soft photon is calculated separately, only the ultra-soft scale remains
in (\ref{generalpropagator}). We represent this separation graphically
in Fig.[4] where a general transverse photon (on the lhs) is represented
by a slanted wavy line and, on the rhs, the soft photon contribution is
represented by a vertical (instantaneous) wavy line and the ultra-soft 
contribution is represented by a slanted,  broken,  wavy line.  To get
the ultra-soft propagator, we must therefore subtract from the general
propagator the expression corresponding to the soft photon propagator
which we have seen in (\ref{softphoton}) to be $- {\cal P}_{ij} / \vec k^2$
(notice, however, that we would operate this subtraction in a diagram
like Fig.[9] where there is no corresponding soft photon contribution):
\ba
 -  {{\cal P}_{ij} \over 2 k} && \biggl( {1 \over 
\gamma^2/(2 \mu)  +
 (\vec p - \vec k)^2/(2 m_1) 
+
\vec p^2/(2 m_2) +k } \nonumber \\ &&~~~~~~~~
~+~ {1 \over  
\gamma^2/( 2 \mu)  +
\vec p^2/(2 m_1) + (\vec p - \vec k)^2/(2 m_2) +k }  - {2 \over k}
\biggr).  \label{gee}
\ea
This expression now corresponds to the propagator of an ultra-soft
photon so 
 the scale of $k$ is of order $\mu \alpha^2$.
Recalling that the scale of $p$ is $\simeq \mu \alpha$, we can perform a
Taylor expansion in $k/p \simeq \alpha$. Applying this to
(\ref{gee}) gives
\ba
   \approx
 - {1 \over 2 k} {\cal P}_{ij} && \biggl[ 
    {1 \over\gamma^2/(2 \mu)  + \vec p^2/(2 \mu) + k} -{ 1 \over k} 
+ {\vec p \cdot \vec k / m_1  \over
\biggl(\gamma^2/(2 \mu)  + \vec p^2/(2 \mu) + k \biggr)^2}
\label{first}  \\
&& + { - \vec k^2/ (2 m_1) \over
\biggl(\gamma^2/(2 \mu)  + \vec p^2/(2 \mu) + k \biggr)^2}  
 + {
 ( \vec p \cdot \vec k)^2/m_1^2 
 \over
\biggl(\gamma^2/(2 \mu)  + \vec p^2/(2 \mu) + k \biggr)^3} \label{second}
\\
 && +
  {
   - \vec k^2 \vec p \cdot \vec k /m_1^2
 \over
\biggl(\gamma^2/(2 \mu)  + \vec p^2/(2 \mu) + k \biggr)^3}
 +
 { (\vec k \cdot \vec p)^3/m_1^3 +   \over 
\biggl(\gamma^2/(2 \mu)  + \vec p^2/(2 \mu) + k \biggr)^4} + \ldots
\biggr]  \nonumber \\  && \quad \quad \quad
 + {\rm~ same~with~}m_1 \leftrightarrow m_2 
\label{third}
\ea
where the first line contain the zeroth order term plus the first order
one (the $\vec p \cdot \vec k$ term), the second line contains the
second order contribution and so on.

Since the expansion is in $k/p$, we expect that each power of $\vec k$
appearing in the numerator will be associated with a power of $\alpha$
with respect to the zeroth order term of the Taylor expansion
(the first term in (\ref{first})).
 We will show this explicitly for a few terms.
 
Consider first the zeroth order 
propagator. It
contains two inverse powers of $k$, which scales like $\mu \alpha^2$, so
that it contributes to the counting rules by a factor $1/ (m^2
\alpha^4)$ (we won't distinguish between $m_1, m_2$ and $\mu$ to discuss
the counting rules).  Of course, in an actual diagram, other factors
will enter to make the overall  $\alpha $ contribution of the diagram
positive; here we are just interested in  the relative contribution of
the terms in the Taylor expansion.

Now consider the  first order correction (the second term  of
(\ref{first})). 
 The numerator $\vec k
\cdot \vec p / m $ scales like $m \alpha^2 \times m \alpha / m = m
\alpha^3$ and the denominator contains $3$ factors of $k$ so it scales
like $( m \alpha^2)^3$. Therefore, the first order  propagator
scales like
\ba
m \alpha^3 / (m^3 \alpha^6) = 1/(m^2 \alpha^3)
\ea
 which is
one power of $\alpha$ times the zeroth order propagator.  The $\vec
k^2/m$
term in  the second order Taylor propagator (\ref{second})
scales like
\ba
(m \alpha^2)^2/(m \times m^3 \alpha^6) = 1/(m \alpha)^2
\ea
 which is down by two
powers of $\alpha$ with respect to  the zeroth order 
propagator.
 It is a simple matter to verify that the other term of
(\ref{second})
 also contribute with a factor of $\alpha^2$ with respect to the
lowest order contribution, and the terms of (\ref{third})
contribute with a
factor  $\alpha^3$, {\it etc}. 

In an actual diagram, the  Taylor expansion  must of course be carried
on the  whole
diagram. As an illustration, we   expand the complete integrand
corresponding to Fig.[3(a)], sandwiched between ground state
wavefunctions:
\ba
 {8 \sqrt{ \pi \gamma^5} \over (\vec p^2 + \gamma^2)^2} 
 &&{q_1 q_2 \over 4 m_1 m_2} ~
(2  p_i -  k_i) \, (k_j - 2  p_j )  
{{\cal P}_{ij}
\over 2 k} ~  \nonumber \\  && 
\biggl({ 1
 \over -\gamma^2/(2 \mu) - (\vec p - \vec k)^2 / (2 m_1)
- \vec p^2/(2 m_2) -k } +{1 \over k}  \biggr)
~   { 8 \sqrt{ \pi \gamma^5} \over  ( (\vec p - \vec k)^2
+ \gamma^2 )^2 } \label{fulltaylor} \nonumber \\
=~
{\bigl( 8 \sqrt{ \pi \gamma^5} \bigr)^2 \over (\vec p^2 +
\gamma^2)^4} ~&&{q_1 q_2 \over 4 m_1 m_2} {{\cal P}_{ij} \over 2 k} {4
 p_i  p_j \,
  \vec p \cdot \vec k  / m_1\over ( - \gamma^2 / (2 \mu) - \vec p^2 / (2
\mu) -k )^2 }~ \nonumber \\  &+&~
{\bigl( 8 \sqrt{ \pi \gamma^5} \bigr)^2 \over (\vec p^2 +
\gamma^2)^4} ~{q_1 q_2 \over 4 m_1 m_2} {{\cal P}_{ij} \over 2 k} 
\biggl(
{1 \over  (-
\gamma^2 / (2 \mu) - \vec p^2 / (2 \mu) -k ) } 
 + {1 \over k} \biggr ) 
\nonumber \\&& ~~~~~~~~~~~~~~~~~~~ \times
~\biggl[ 
-4  p_i  p_j + 2   k_i   p_j + 2   p_i   k_j 
~ -{ 16  p_i  p_j \,
 \vec p \cdot \vec k \over (\vec p^2 + \gamma^2)} +
\dots \biggr] \label{taylordiag}
\ea
Again, one can easily verify that each power of $\vec k$ in the
numerator is associated with an extra factor of $\alpha$.

Notice that the spin-spin diagram with an ultra-soft photon, Fig.[8(b)],
contains at least two powers of $\vec k$ since the NRQED Feynman rule
for the Fermi vertex is proportional to $\vec k$; in other words, the
first non-vanishing contribution comes from the second order term in the
Taylor expansion. Therefore, the lowest
order contribution of the ultra-soft spin-spin exchange is suppressed by
two powers of $\alpha$ with respect to the corresponding  dipole-dipole
 exchange (this is due to the fact that the
Fermi interaction involves the $\vec B$ field). This is very different 
from the corresponding soft photon diagrams which both contribute to the
same order. The difference, again, is that only the factors of $e$ and
$1/m$ enter in the soft photon counting rules whereas factors of the
photon momentum $\vec k$ matter in the ultra-soft counting rules.

Clearly, the fact that one power of $\alpha$ is generated by each term
in the Taylor expansion will prove crucial in writing down the
counting rules of this new, Taylor expanded, version of NRQED.
However, before doing so, we now want to show that the Taylor expansion
we just carried is equivalent to a multipole expansion of the NRQED
vertices.

\subsection{Connection with the multipole expansion}

As an example, consider the $ - q \psi^\dagger 
  (\vec p \cdot \vec A + \vec A \cdot \vec p)/(2m) \psi $
interaction contained in the term  $ - \psi^\dagger
\vec
D^2/(2m) \psi$ in the hamiltonian. 
 To obtain the  NRQED Feynman rule we first replace the
fields by plane waves:
\ba
 + q i \bigl( { e^{-i \vec p' \cdot \vec r} \nabla \cdot \vec \epsilonv 
 \biggl(e^{-i \vec k \cdot \vec
r} e^{i \vec p \cdot \vec r} \biggr)  \over 2 m}  + {
e^{-i \vec p' \cdot \vec r} \biggl(e^{-i \vec k \cdot \vec
r} \vec \epsilonv \cdot \nabla  e^{i \vec p \cdot \vec r} \biggr)
\over 2 m}   \bigr) \label{pdotA}
\ea
where $\vec p, \vec p' $ are respectively the three momenta of
the fermion line before and after the interaction, and $\vec k$ is the
photon three-momentum; $\epsilonv$ is the photon polarization. Applying
the derivatives, we get 
\ba
 { q \over 2 m }
  \biggl(  k_i -  2  p_i \biggr) e^{-i (\vec p' + \vec k - \vec p )
\cdot \vec r}  
. \label{Feynmanrule}
\ea
The exponential leads, as usual, to the conservation of three-momentum
$\vec p' = \vec p - \vec k$ (here we considered a photon being emitted).
Using this to write $-2 \vec p = - \vec p -\vec p' - \vec k$ and
discarding  all factors associated with the external fields,  we obtain 
the Feynman rule  
\ba
  -  q  { p_i +  p_i' \over 2m}  .
\ea
The rule is obviously unchanged if we consider an absorbed photon.
Now we consider a multipole expansion of this vertex, {\it i.e.}
we expand the photon field
\ba
e^{-i \vec k \cdot \vec r} = 1 -i \vec k \cdot \vec r +{1 \over 2} ( -
i\vec
k \cdot \vec r)^2 + \ldots
\ea
In the following, we will use the notation $e^{-i \vec k \cdot \vec r}
= {\rm zeroth~ order} + {\rm first~ order} + \ldots$ to label the terms
in the multipole expansion.
As usual, this expansion makes sense only if $kr \ll 1$. The size of $r$
is set by the bound state to be of order the Bohr radius $r \simeq
 1 / \gamma$. For ultra-soft photons, we have $k \simeq \gamma^2/\mu$
so that $k r \simeq \alpha$ and the multipole expansion is valid. Of
course, it would be nonsensical to use it for soft photons. Also,
the multipole expansion is clearly the same as the Taylor expansion
performed above since the scale of $r \simeq 1/ p$.

We can easily find the rule for the new vertex. Using the first term
of the multipole expansion, $e^{-i \vec k \cdot \vec r } = 1$ 
 (corresponding to an
E1 transition) in (\ref{pdotA}), we obtain
\ba
- {     p_i  \over m } ~ e^{-i (\vec p'  - \vec p )
\cdot \vec r}  
\ea
where now the exponential leads to the condition 
 $\vec p' = \vec p$, {\it i.e.}  three
momentum is {\it not } conserved at the vertices when the multipole
expansion is used. This can, however, still be used to write the rule
for the vertex as  before, {\it i.e.}
\ba
   -q  { p_i +  p_i' \over 2m} . \label{dipole}
\ea
Even though the rule for the vertex is the same as before,
 the condition
 $\vec p' = \vec p$ simplifies greatly  the expression for diagrams containing
ultra-soft photons and, in particular, the propagator.
To see this,  we  first  go back to the time
ordered photon-fermion pair propagator (\ref{gee}).
   We now use in addition the
fact that the fermion momenta at  the vertices
are unchanged by the emission
or absorption of the photon to write (\ref{gee}) as 
\ba
 - {{\cal P}_{ij} \over 2 k}   &&   \biggl( {1 \over 
\gamma^2/(2 \mu )  +
\vec p^2/(2 \mu)  +k }
~+~ {1 \over  
\gamma^2/( \mu)  +
\vec p^2/(2 \mu)  +k }  - {2 \over k }   \biggr)
 \nonumber 
\\ &=&- {{\cal P}_{ij} \over k}   \biggl( {1 \over
\gamma^2/(2 \mu )  +
\vec p^2/(2 \mu)  +k } - {1 \over k } 
 \biggr)
 \label{prop3}
\ea
instead of the form (\ref{gee}) 
which was obtained by using $\vec p' = \vec
p - \vec k$. 
  In  (\ref{prop3}), the scale of $k$ is set either by
$\gamma^2/(2 \mu) \simeq \mu \alpha^2$ or $\vec p^2 / (2\mu)$, but   since
$\vec p$ is a fermion three-momentum it is of order $\gamma$, 
we get in either case 
   $ k \simeq \mu \alpha^2$.
This shows explicitly that the multipole expansion has permitted us to
isolate the 
ultra-soft scale.

To obtain the  higher order terms  in the multipole expansion, one
provides a factor 
 $(\pm \vec k \cdot
\vec \nabla_{\vec p})^n/n !$ for each vertex connected to an ultra-soft
photon,  where $n$ is the order of interest in the multipole
expansion, and a plus (minus) sign is used if the photon is absorbed
(emitted). In this expression, the gradient must be taken with respect
to the three-momentum of the fermion line on the {\it right} of the
vertex. To apply these rules, it is therefore necessary to distinguish
between the momentum of the fermion before and after the interaction,
even though we have to  set them equal in the end.

To illustrate this, we will evaluate the first few multipole corrections
to the ultra-soft photon propagator.
 Since,  as noted above, 
 one must
 distinguishes the  momenta of each fermion and the momenta
before and after the interaction, we will use the momenta as labeled
in Fig.[6], with the understanding that one must set 
\ba
\vec p_1 = \vec
p_1' = -\vec p_2 = - \vec p_2' = \vec p \label{replace}
\ea
 after carrying out the
derivatives. Taking this into account, 
the intermediate state propagator in Fig.[6(a)]  is 
\ba
 - { 1 \over 2 k} {\cal P}_{ij}
\bigl( {1 \over k + \gamma_2/(2\mu) + \vec p_1^{' \, 2}/ (2 m_1) + \vec p_2^2/
 (2 m_2) }  -{1 \over k}  \bigr) \label{prop1}
\ea
and the propagator of Fig.[6(b)] is
\ba
 - { 1 \over 2 k} {\cal P}_{ij}
\bigl( {1 \over k + \gamma_2/(2\mu) + \vec p_1^2/ (2 m_1) + \vec
p_2^{' \, 2} / (2 m_2) } -{1 \over k}   \bigr) . \label{prop2}
\ea

If we consider Fig.[6(a)], then we only have to consider the multipole 
expansion of the vertex on the upper line since the other vertex will
not act on the intermediate state propagator. We therefore apply, as
we did above, the operator $- \vec k \cdot \vec \nabla_{\vec p_1'}$
on (\ref{prop1}) to obtain
\ba
  { { \cal P}_{ij} \over 2 k }    {  - \vec k \cdot \vec p /
m_1 \over \bigl( k +
\gamma_2/(2\mu) + \vec p^2/ (2 \mu) 
\bigr)^2 }  . \label{MQED1}
\ea
In the case of Fig.[6(b)], we apply the operator $ i \vec k \cdot \vec
\nabla_{\vec p_2'}$ on (\ref{prop2}) with, for result, (recall that we
replace $\vec p_2$ by $ - \vec p$ after differentiating) 
\ba
 - { {\cal P}_{ij} \over 2 k } 
    { \vec k \cdot \vec p / m_2 \over \bigl( k +
\gamma^2/(2\mu) + \vec p^2/ (2 \mu) 
\bigr)^2 } . \label{MQED2}
\ea
 As expected, this is the same as (\ref{MQED1})
with $m_2$ replaced by $m_1$.
  The sum of (\ref{MQED1}) and (\ref{MQED2})  is the
result of the first order  term of the multipole expansion.  To be more
precise, this is the result obtained from considering the first order
term in the multipole expansion of either vertex.

The result of the second order multipole can easily be calculated
in a similar way. We apply the operator $(- \vec k \cdot \vec
\nabla_{\vec p_1'})^2/2 $ to (\ref{prop1}) and  $( \vec k \cdot \vec
\nabla_{\vec p_2'})^2/2 $ to (\ref{prop2}) to obtain
\ba
&&  
 {{\cal P}_{ij} \over 2k }  \biggl({ \vec k^2/ (2 m_1)  \over \bigl( k +
\gamma^2/(2 \mu) + \vec p^2/(2 \mu) \bigr)^2}
- { (\vec k \cdot \vec p )^2 /m_1^2  \over \bigl( k +
\gamma^2/(2 \mu) + \vec p^2/(2 \mu) \bigr)^3}  + {\rm same~with~}m_2
\rightarrow m_1 \biggr)
. \label{secondorder}
\ea
 These expressions correspond to keeping  the $n=2$
multipole term on either of the vertices plus   the first order
term on both vertex, all of  which  contribute to the same order in $\alpha$,
as we will discuss in the next section.

 We also give the third order result:
\ba
{{\cal P}_{ij} \over 2 k} \biggl( 
 {  \vec k^2 \vec k \cdot \vec p /  m_1^2  \over \bigl( k +
\gamma^2/(2 \mu) + \vec p^2/(2 \mu) \bigr)^3}
- { (\vec k \cdot \vec p )^3 /m_1^3  \over \bigl( k +
\gamma^2/(2 \mu) + \vec p^2/(2 \mu) \bigr)^4}
~+~{\rm same~with~}m_2 \rightarrow m_1 \biggr)
. \label{thirdorder}
\ea

We have recovered the expressions obtained from the Taylor expansion, 
Eqs.(\ref{first}, \ref{second} and \ref{third}).
 This is not surprising since the Taylor expansion of a
function $f(x+a)$ around $x=0$  can be written
as 
\ba
f(x+a)_{x \simeq 0}~=~ e^{x\, { d \over da}} \, f(a)
\ea
and this is what the multipole expansion accomplishes.

In an actual calculation, the multipole expansion must of course
be carried on the whole diagram. This is slightly more complex because
the wavefunctions must also be written in a way that distinguishes
the momenta on each fermion line. 
To illustrate this, we consider again the bound state diagram 
corresponding to Fig.[6(a)] and work out the expression in first order 
of the  multipole expansion.
We again use the ground state
wavefunction (\ref{wave}) for the external states.
 Taking this into account, the integrand corresponding to
Fig.[6(a)] is given by 
\ba
 &&{ 8 \sqrt{ \pi \gamma^5} \over \mu^2 ( \vec p_1^2
 /m_1 + \vec p_2^2 / m_2 + \gamma^2/ \mu)^2 }
{q_1 q_2 \over 4 m_1 m_2} 
(\vec p_1 + \vec p_1')_i \, (\vec p_2 + \vec p_2')_j  
{{\cal P}_{ij}
\over 2 k} ~  \nonumber \\&& \bigl({ 1
 \over -\gamma^2/(2 \mu) - \vec p_1^{'\,2} / (2 m_1)
- \vec p_2^2/(2 m_2) -k } +{1 \over k}  \bigr) 
   { 8 \sqrt{ \pi \gamma^5} \over \mu^2 ( \vec p_1^{'
\, 2} /m_1 + \vec p_2^{' \, 2} / m_2 + \gamma^2/ \mu)^2 } \label{papa}
\ea
The contribution of the zeroth order in the multipole expansion is
obtained by simply using the relations (\ref{replace}) directly in 
(\ref{papa}). 
 The
contribution of the first order multipole expansion is then obtained by
applying on this expression the operator $- \vec k \cdot \vec
\nabla_{\vec p_1'} $, which is associated with the vertex on the left
in Fig.[6(a)]
 plus the operator $  \vec k \cdot \vec
\nabla_{\vec p_2'}$ for the second vertex, and then reexpressing the
vectors in terms of $\vec p$ using (\ref{replace}). The result is
\ba
&&{q_1 q_2 \over 4 m_1 m_2}~
{ 8  \sqrt{ \pi \gamma^5})^2 \over (\gamma^2 + \vec p^2)^4 }~
{ { \cal P}_{ij} \over 2 k} ~ \biggl(
  { 4 \vec p_i \vec p_j \vec k \cdot \vec p / m_1 \over 
(-\gamma^2/(2 \mu) - \vec p^2 / (
2 \mu) - k)^2 }  \biggr)    \nonumber \\ &+&~
{q_1 q_2 \over 4 m_1 m_2}~
{ 8  \sqrt{ \pi \gamma^5})^2 \over (\gamma^2 + \vec p^2)^4 }~
{ { \cal P}_{ij} \over 2 k} ~ \biggl(
  { 1 \over -\gamma^2/(2 \mu) - \vec p^2 / (
2 \mu) - k } + {1 \over k} \biggr)    \nonumber \\&&~~~~~~~~~\quad
\quad \quad \quad  \times 
\biggl( 2 \vec k_i
 \vec p_j + 2 \vec k_j \vec p_i  
 - 16  \vec p_i \vec p _j \, { \vec k \cdot \vec p  \over
 ( \vec p^2 +
\gamma^2) } \biggr) . \label{bobo}
\ea
This is, as expected,
 equal to the expression obtained from the first order Taylor
expansion (\ref{taylordiag}).
 A similar calculation  for
 Fig[6(b)]
gives the same   result   as (\ref{bobo}) with $ m_1 \leftrightarrow
m_2$.

Notice that the zeroth order term in the multipole expansion is obtained
by setting $\vec p' = \vec p$ in the NQRED vertices. In the case of the
Fermi vertex, this gives zero since the NRQED Feynman rule is
proportional to $\vec p' - \vec p = \vec k$. This means that the first
nonzero contribution is of the first order in the multipole expansion.
Higher order terms are obtained as above {\it i.e.} by considering the
corresponding factor of 
 $(\pm \vec k \cdot
\vec \nabla_{\vec p})^n/n !$.

Even though we have simply recovered the expressions obtained by
performing a simple Taylor expansion, there is one important reward for
doing so: one can use directly the Wigner-Eckart theorem and the
familiar selection rules derived in quantum mechanics for each
interaction generated by the Taylor expansion. This has consequences in
decays of positronium, and in nonrelativistic QCD bound states
\cite{NRQCD}. This will be explored elsewhere \cite{Pat}.

To summarize, we have seen that, starting from NRQED, separating the
soft and ultra-soft scales and applying a multipole expansion  (or
Taylor expanding) the vertices connected to ultra-soft photons
generates a new effective theory with its own set of Feynman rules.
This theory, which we  will call  ``MQED" (for ``Multipole QED")
has the advantage of generating bound state diagrams that contribute to
a unique order in $\alpha$. In the last section, we will derive the MQED
Feynman rules and show some applications of the  counting rules.

\section{MQED counting rules}

We can now easily extend the counting rules to include diagrams
containing ultra-soft photons. The concept of potentials is not
well-defined, however, when ultra-soft photons are present, so we first
rewrite the soft counting rule (\ref{lambda})  as a sum over vertices
instead of a sum over potentials:
\ba
\zeta  ({\rm soft~photons})= \sum_{j=1}^{N_{{\cal V}}} \bigl(
  d_j +  n_j + l_j \bigr) +1- N_{TOP}  \label{softcounting}
\ea
where ${\cal V}$ is the total number of vertices contained in the
diagram. For a diagram containing ultra-soft photons, this rule must be
changed to
\ba
\zeta = \sum_{j=1}^{N_{{\cal V}}} \bigl(
  d_j +  n_j + l_j \bigr) +1- N_{TOP} + 2 N_{
\gamma} +   {\cal M}_{ultra-soft}   \label{countingrule}
\ea
where  $ N_{
\gamma}$ is the number of ultra-soft photons in the diagram.
The last term, ${\cal M}_{ultra-soft}$ can be expressed in two different
ways, depending on whether one uses a Taylor expansion of the diagram or
a multipole expansion of the vertices.
In the first case, ${\cal M}_{ultra-soft}$ is simply the power of $\vec
k$ appearing in the numerator. In the second case, ${\cal
M}_{ultra-soft}$  can be written as
\ba
{\cal M}_{ultra-soft}= \sum_{i} {\cal M}_i
\ea
where the sum is over the vertices connected to ultra-soft photon
and ${\cal M}_i$ is the order in the multipole expansion to which the
$i^{th}$ vertex has been expanded.

 Eq.({\ref{countingrule})  gives the
order,  in powers of $\alpha$, at which an arbitrary MQED diagram will
contribute. The dependence on arbitrary masses  is unchanged
by the presence of ultra-soft photons and is therefore still given by 
(\ref{order1}). The charge dependence, however, is different when there
are ultra-soft photons because the ultra-soft scale is $ \simeq
\gamma^2/ \mu \simeq \mu  Z^2 \alpha^2 $ so that  the  $Z$ dependence
 is different then  in
the soft scale $ \gamma = Z \mu \alpha$. The expression for the charge
dependence must then be changed from  (\ref{eta}) to
\ba
\eta = 1   - N_{TOP} + \sum_{i=1}^{{\cal V}} \bigl(  d_i + a_i \bigr)
 + 2 N_{
\gamma} + \sum_{u-s {\cal V}_i}  {\cal M}_i 
\label{eta2}
\ea
where, again, the second sum is over the vertices connected to
ultra-soft photons only.

Our final result is therefore that an arbitrary  MQED diagram 
 will contribute to order
\ba
{\cal O} =   {\mu^{\kappa+ \rho +1 } \over m_1^\kappa m_2^\rho} Z^\eta
 \alpha^\zeta  \label{finalresult}
\ea
with $\zeta$ defined in (\ref{countingrule}), $\eta$ defined in
(\ref{eta2}) and $\kappa$ and $\rho$ defined in (\ref{kappa}).

We now give a few examples of the use of (\ref{finalresult}). As a
first example consider the interaction Fig.[7] in hydrogen,
 where the ultra-soft photon is connected
to an electron line. In this diagram, $d_1 = d_2 = 1$ (there is
one factor of $1/m$ on each vertex), $n_1 = n_2 = 1/2$ (a factor $e$ on
each vertex), $l_1= l_2 =0$ (the $\vec p \cdot \vec A$ interaction does
not get renormalized), $N_{TOP}= 1$, $N_\gamma =1$ and, if the zeroth
order in the multipole expansion (or in the Taylor expansion) is used,
 ${\cal M}_1 = {\cal M}_2 =0$. This leads to a contribution of order
$\alpha^5$. The mass dependence is found to be $\mu^3 / m_e^2$ and the 
$Z$ dependence is, from (\ref{eta2}),  $Z^4$. This diagram therefore
contributes to order
\ba
{ \mu ^3 Z^4 \over m_e^2} \alpha^5 .
\ea
In fact, this result is enhanced by a logarithm $\ln (Z \alpha)$ and 
contributes to the Lamb shift.

Consider now  Fig.[8(a)] in positronium so that $Z=1$ and $m_1= m_2
=m_e$. In this diagram,    the transverse
photon is soft (it is represented by a vertical line). We can therefore
use (\ref{softcounting}). One has $n_1 =n_2 =1/2$ and $d_1 = d_2 = 1$.
If the tree level expressions are used for the coefficients, then this
diagram contributes to order $m_e \alpha^4$. The same diagram will
contribute to higher order in $\alpha$ if the loop corrections to the
coefficients of the Fermi vertices are considered (the one-loop
correction being, from (\ref{renor}), $\alpha/2 \pi$).

As a final example, consider  Fig.[8(b)].
 Here the photon is ultra-soft.  As mentioned previously, the first
nonvanishing contribution from this diagram   contains two factors of
$\vec k$ (one from each spin vertex) so that ${\cal M}_{ultra-soft}$ in
(\ref{countingrule}) is at least equal to two.  $N_{TOP}= 1$,
$N_{\gamma} =1 $ and the other coefficients are as in Fig.[8(a)], if the
tree level coefficients are used. One the finds that this diagram will
contribute to order $m_e \alpha^7$.

\acknowledgments

 I have  benefited from many useful conversations with
Peter Lepage who suggested first the idea of applying the multipole
expansion to NRQED. 
I  also want to
 thank S.M. Zebarjad for several very useful comments and for drawing
the figures.
This work was supported by an NSERC (Canada) postdoctoral fellowship, and
by
les fonds FCAR du Qu\'ebec.

\begin{figure}[t]
\centerline{\epsfysize=8.0true in \epsfxsize=4.2 true in
\epsfbox{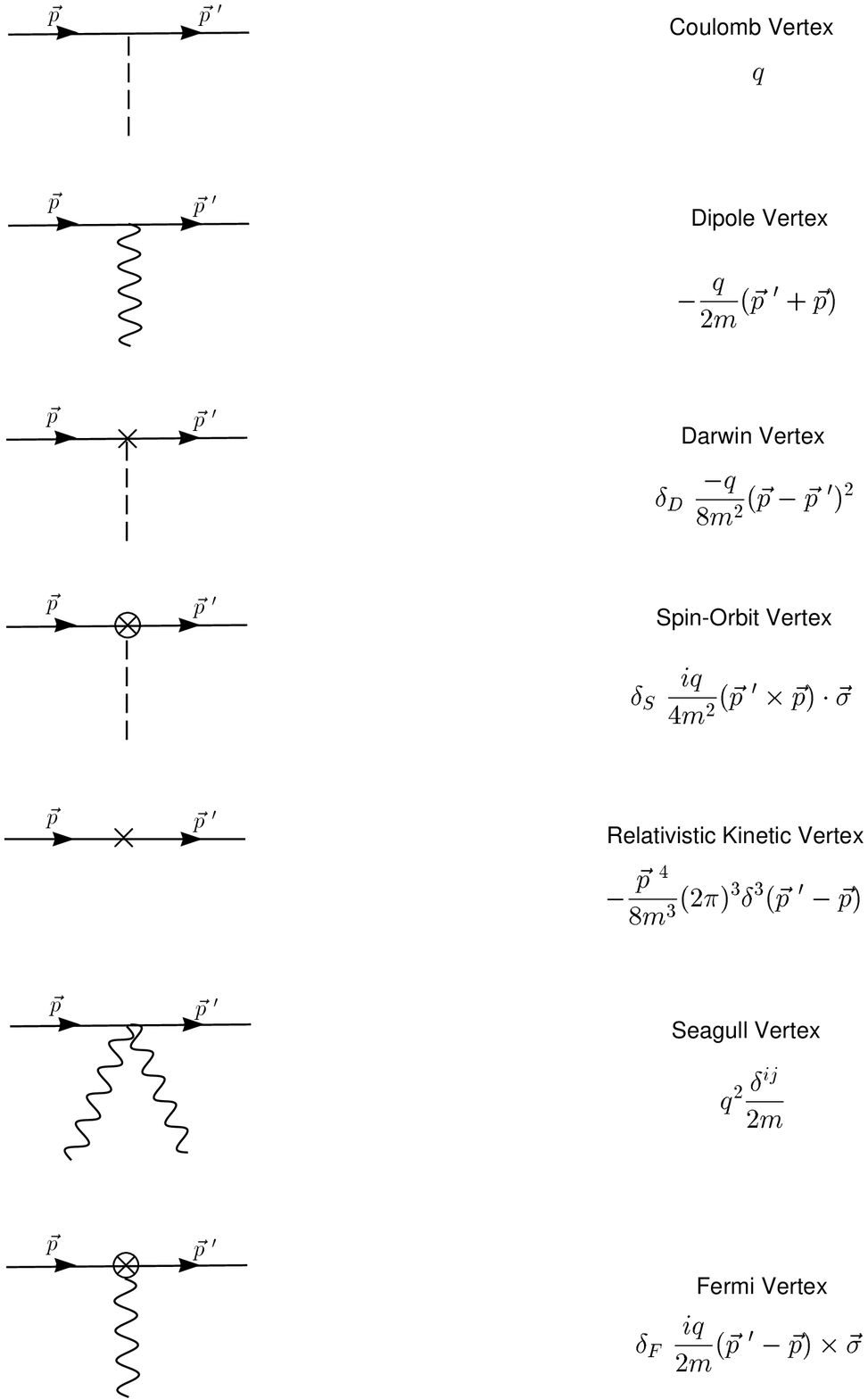}}
\vspace{3ex}
\caption{NRQED Feynman rules  \label{feynmanrules}
}
\vspace{3ex}
\end{figure}
\vfill\eject
\begin{figure}[t]
\centerline{Feynman rules (continued)}
\centerline{\epsfysize=5.7true in \epsfxsize=4.2 true in
\epsfbox{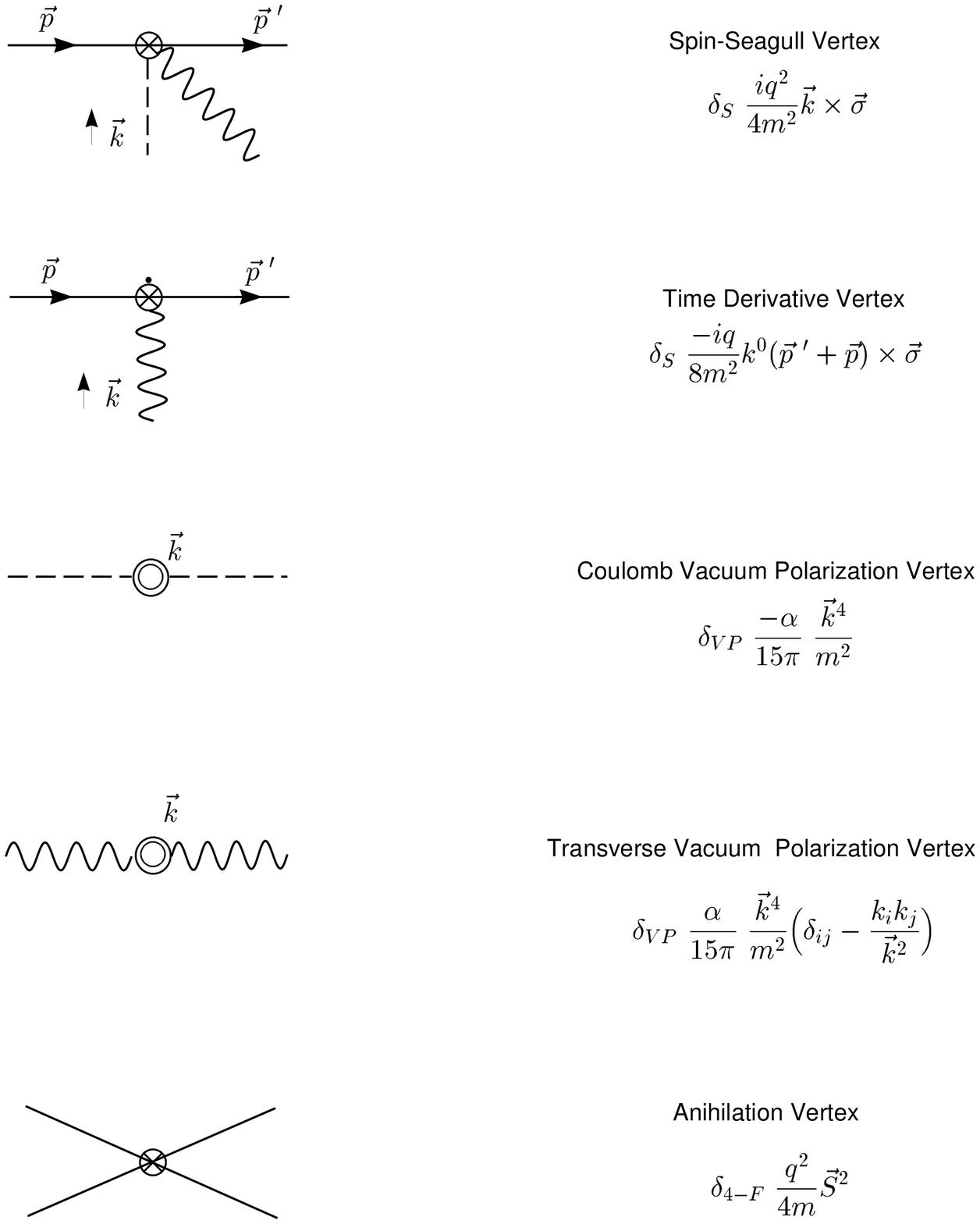}}
\vspace{3ex}
\vspace{3ex}
\end{figure}
\vfill\eject
\begin{figure}[t]
\centerline{Feynman rules (end)}
\centerline{$~$}
\centerline{$~$}
\centerline{\epsfysize=3.4true in \epsfxsize=5.5 true in
\epsfbox{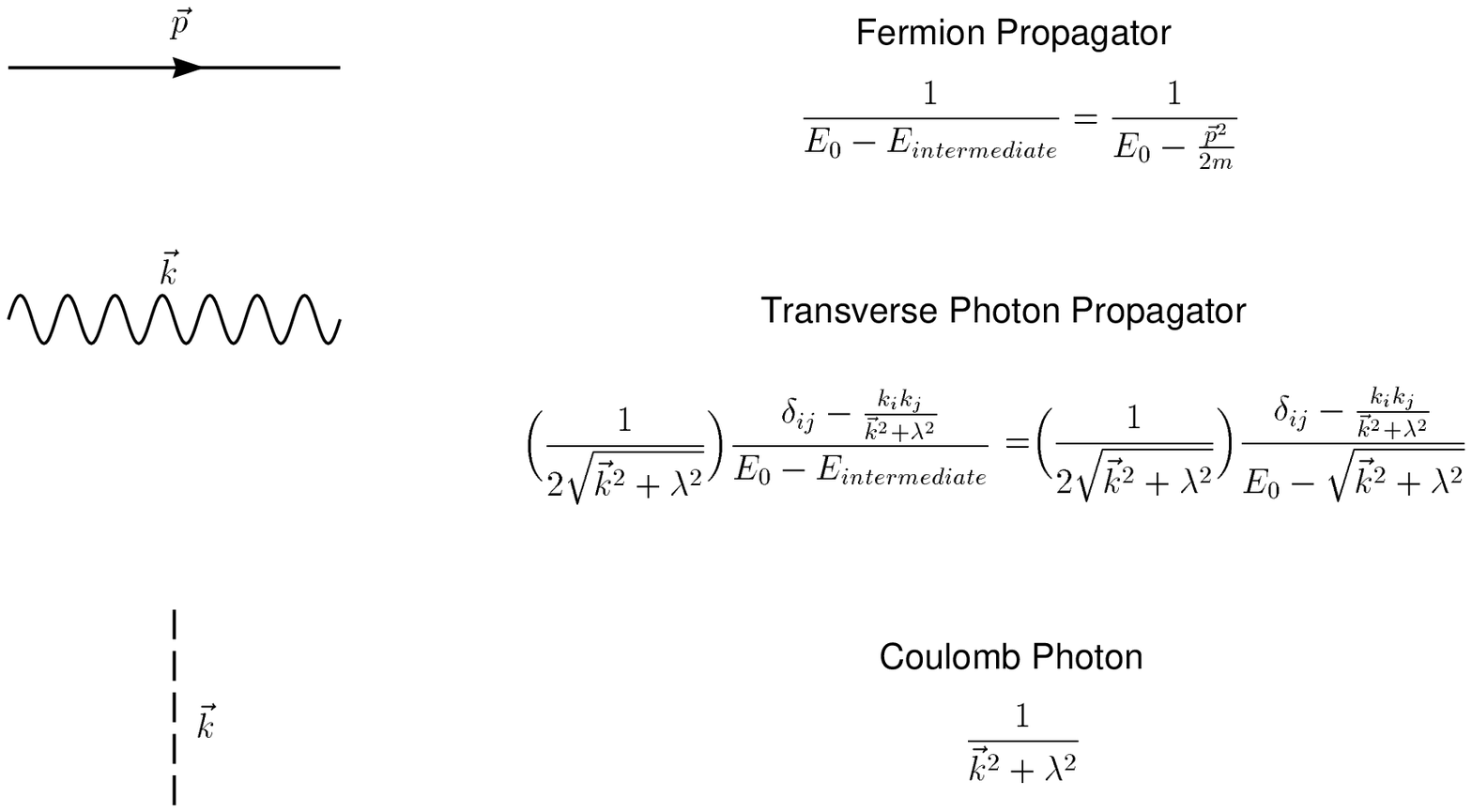}}
\vspace{2true in}
\vspace{9ex}
\end{figure}
\vfill\eject
\begin{figure}[t]
\centerline{$~$}
\centerline{$~$}
\centerline{\epsfysize=2.4true in \epsfxsize=5.5 in
\epsfbox{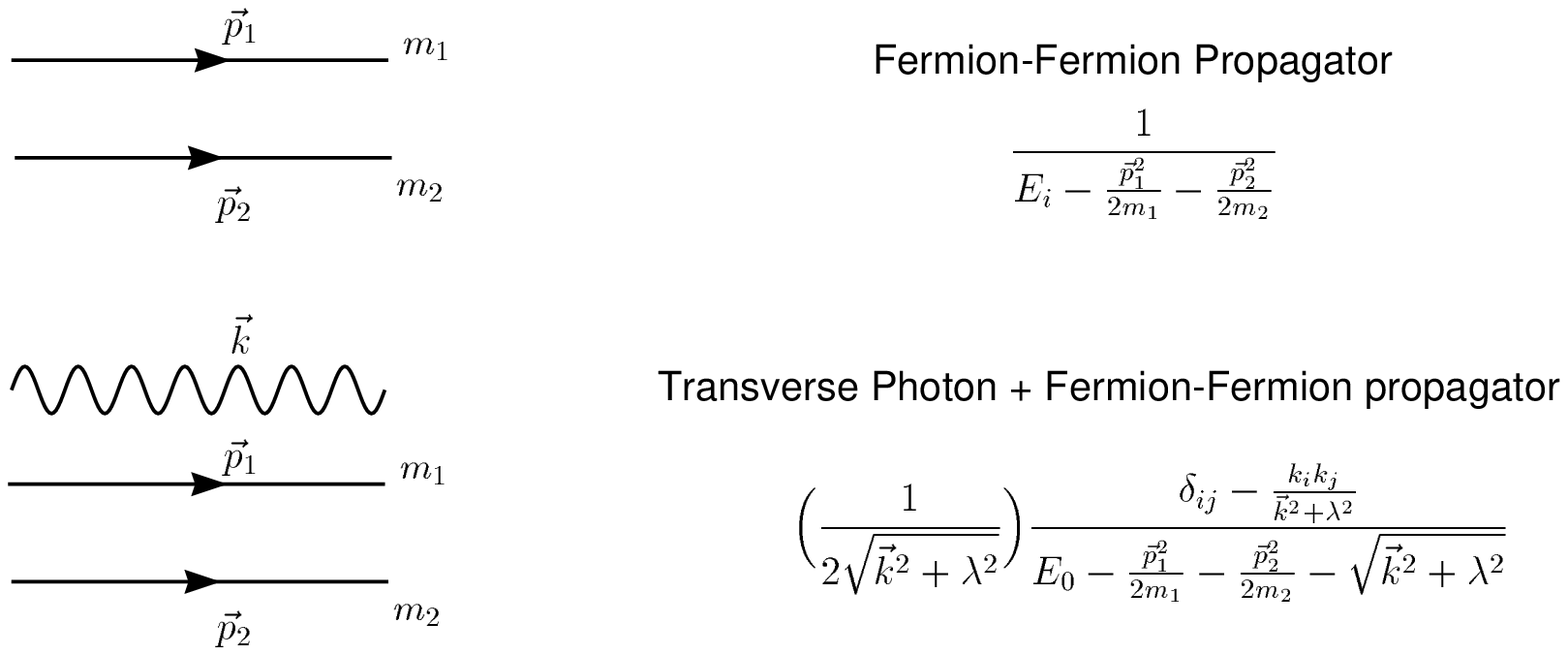}}
\vspace{2true in}
\vspace{9ex}
\caption{Time-ordered propagators for two fermions or two fermions
plus one transverse photon.}
\end{figure}
\vfill\eject
\begin{figure}[t]
\centerline{\epsfysize=2.2true in \epsfxsize=5.2 true in
\epsfbox{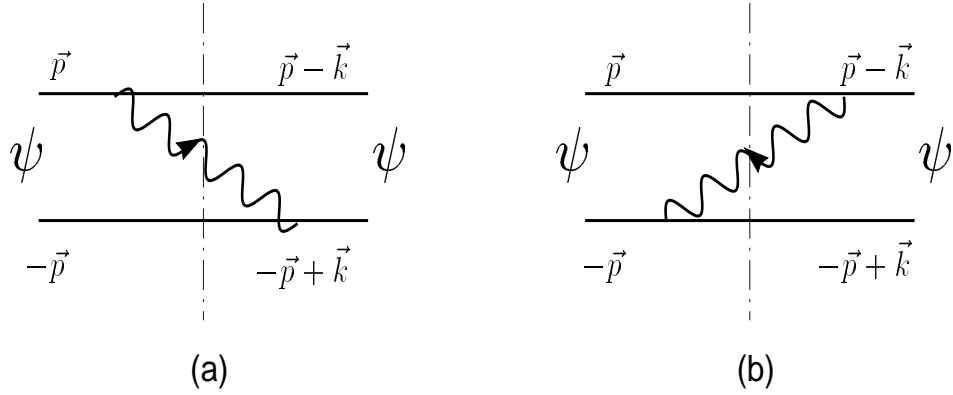}}
\caption{The two time-ordered diagrams corresponding to the exchange of
a transverse photon (the vertical lines indicate the  intermediate states
used for the time-ordered propagators).}
\vspace{3ex}
\vspace{2 true in }
\end{figure}
\begin{figure}[t]
\centerline{\epsfysize=1.2true in \epsfxsize=5.4 true in
\epsfbox{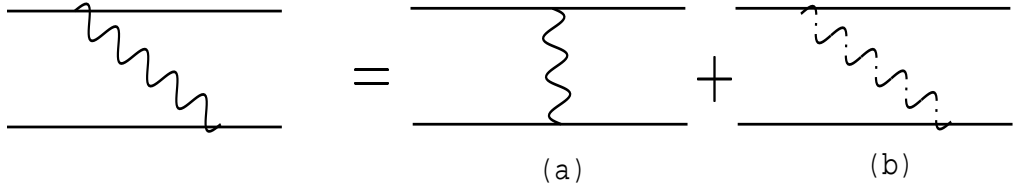}}
\caption{Separation of a transverse photon into a soft, instantaneous
contribution (represented by a vertical line) and an ultra-soft
propagator (represented by the broken wavy line).}
\vspace{3ex}
\vspace{3ex}
\end{figure}
\vfill\eject
\begin{figure}[t]
\centerline{\epsfysize=2.2true in \epsfxsize=4.3 true in
\epsfbox{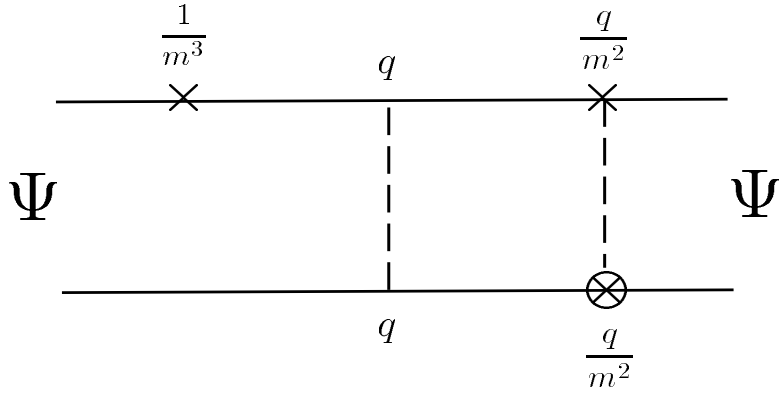}}
\caption{ Generic bound state potential; the dependence on the charges
and on the masses of each vertex is indicated.}
\vspace{3ex}
\vspace{3ex}
\end{figure}
\begin{figure}[t]
\centerline{\epsfysize=2.2true in \epsfxsize=6.2 true in
\epsfbox{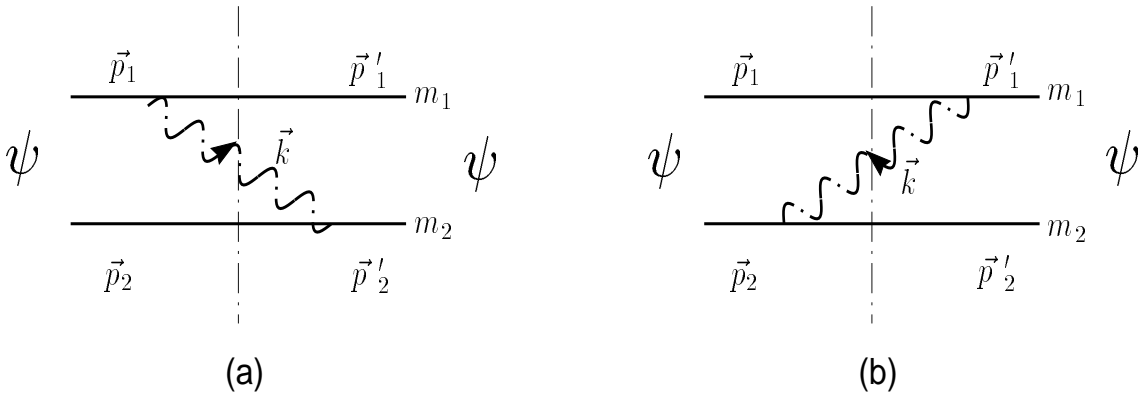}}
\caption{ The two time-ordered diagrams corresponding to a transverse
photon exchange with the routing necessary to apply the multipole
expansion.}
\vspace{3ex}
\vspace{3ex}
\end{figure}
\vfill\eject
\begin{figure}[t]
\centerline{\epsfysize=1.3true in \epsfxsize=2.1 true in
\epsfbox{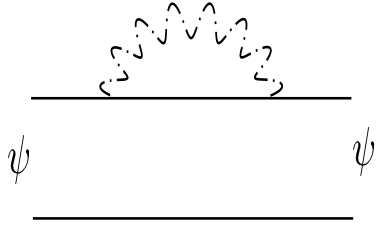}}
\caption{ Self-energy diagram with an ultra-soft photon.
}
\vspace{3ex}
\vspace{3ex}
\end{figure}
\begin{figure}[t]
\centerline{\epsfysize=1.1true in \epsfxsize=4.4 true in
\epsfbox{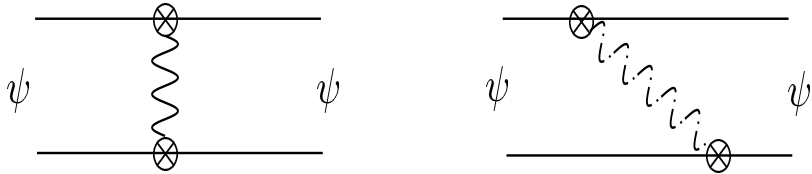}}
\caption{ Spin-spin exchange with a soft photon (vertical line)
and an ultra-soft photon.
}
\vspace{3ex}
\vspace{3ex}
\end{figure}
\begin{figure}[t]
\centerline{\epsfysize=1.6true in \epsfxsize=2.2 true in
\epsfbox{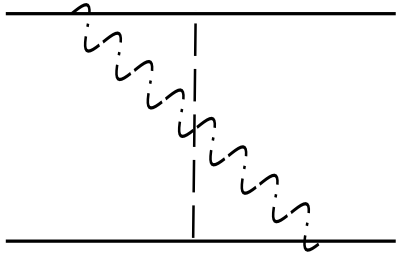}}
\caption{Ultra-soft photon spanning a Coulomb interaction. In such a
diagram, one does not subtract the soft photon propagator from the
intermediate state propagator because there is no corresponding soft
photon diagram.
}
\vspace{3ex}
\vspace{3ex}
\end{figure}
\end{document}